# Process Improvement Archaeology – What led us here and what's next?

Michael Unterkalmsteiner (mun@bth.se) and Tony Gorschek (tgo@bth.se)

Software Engineering Research Lab Sweden

Blekinge Institute of Technology

Karlskrona, Sweden

While in every organization corporate culture and history change over time, intentional efforts to identify performance problems are of particular interest when trying to understand the current state of an organization. The results of past improvement initiatives can shed light on the evolution of an organization, and represent, with the advantage of perfect hindsight, a learning opportunity for future process improvements. We encountered the opportunity to test this premise in an applied research collaboration with the Swedish Transport Administration (STA), the government agency responsible for the planning, implementation and maintenance of long-term rail, road, shipping and aviation infrastructure in Sweden.

The agency was formed in 2010 in order to render the, until then separate road, railway and maritime agencies, more efficient. To achieve the government's goal of increasing productivity and innovation in the construction market, STA targets that by 2018, 50% of the total project volume is realized in the form of design/build contracts, allowing the agency to focus its efforts on core competencies and outsource other activities to suppliers [1]. Also in the software industry, outsourcing of IT services is a common and growing business practice [2]. Design/build contracts are particularly attractive for STA as solution providers can innovate, develop and use components that work as an integrated whole, taking also over responsibilities and risks for activities that were previously carried out by the client [3], i.e. the STA. However, this form of project delivery requires also an increased competence in requirements engineering, communication and management since the design work is outsourced to a solution provider. STA (and its precursors organizations) recognized this need in the early 2000's and invested resources to improve in the area of requirements engineering. Today, STA is very conscious about the importance of requirements engineering and management as part of their overall development processes. Therefore, we decided to perform a Process Improvement Archaeology (PIA) to better understand the current processes and devise new directions for further improvements addressing the currently encountered challenges. Before we look at the PIA steps and its results in more detail, we provide the context and motivation for focusing the investigation on requirements engineering in particular.

## A look over the tea cup's rim

Requirements engineering is a central part in the development of software intensive products, governing planning, effective implementation and product quality [4]. This is even more emphasized in projects where the design and implementation is outsourced to a software supplier [5]. Research and practice in this area has therefore sought to improve and validate requirements engineering concepts in various domains, adapting them to the particular context. A witness to the growing knowledge base on requirements engineering are the

53 systematic literature reviews that were performed between 2006 and 2014, covering nearly 8000 primary studies [6]. Since requirements engineering can be seen as a branch of system engineering [7], our conjecture is that there is a large potential of transferring and applying this knowledge to the construction industry, given the

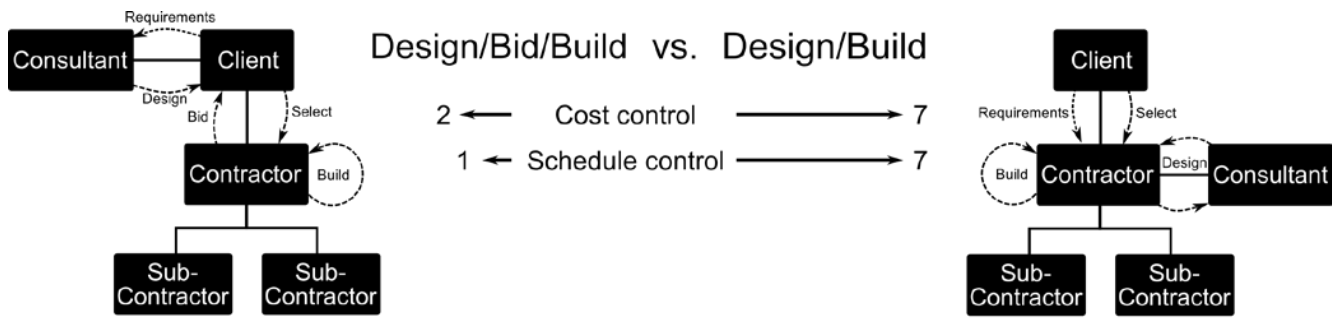

Several studies investigated the impact of the project delivery approach on cost and schedule of project completion. Seven studies determined advantages in both cost and schedule control for the Design/Build approach, while only 2 observed advantages in cost and 1 in schedule control for the Design/Bid/Build approach.
D. R. Hale, P. P. Shrestha, G. E. Gibson, and G. C. Migliaccio, "Empirical comparison of design/build and design/bid/build project delivery methods," Journal of Construction Engineering & Management, vol. 135, no. 7, pp. 579–587, 2009.

*Figure 1: The D/B/B and D/B project delivery paradigm*

growing need for requirements engineering competence, as illustrated next.

The predominant paradigm for the development of large infrastructure projects in the 20$^{th}$ century has been the design/bid/build (D/B/B) project delivery system [8]. In this paradigm (Figure 1, left), the project is separated in a design and a construction phase [9]. The client commissions a consultant to produce bid documents and technical specifications that meet the client's requirements. The bid documents are used to elicit and then select an offer from competing contractors. The winning contractor is commissioned to implement the project according to the specification.

The design/build (D/B) paradigm (Figure 1, right) was the predominant form in pre-industrial times and is now witnessing a renaissance in the wake of downsized in-house project management capabilities and costly disputes between design and construction parties [8]. In this project delivery system, the client deals with a single contractor, responsible for both design and construction services [8]. In contrast to D/B/B, design and construction run in parallel, leading to shorter project delivery times and lower total costs (see Figure 1).

The reason why we are interested in studying construction projects is the central role *requirements* play in the D/B paradigm [9]. In the D/B/B paradigm, requirements can be refined during the design phase of the project where client and consultant define needs and solutions collaboratively [10]. However, the D/B paradigm requires that the client's needs are precisely described such that they can be universally understood and interpreted [10] by all involved stakeholders (client, contractor and consultant). A well-defined and commonly understood scope has been identified as the most important D/B project characteristic [11].

If these observations sound familiar to the Software Engineering ear, it is because high quality requirements are equally important for companies who outsource design, implementation and/or testing of software [12]. Given these parallels in requirements engineering management between construction and software engineering projects, we sought to understand STA's current approach to requirements engineering, and the evolutionary steps in their improvement efforts as they provide justification for the current state and indicate directions for future improvements.

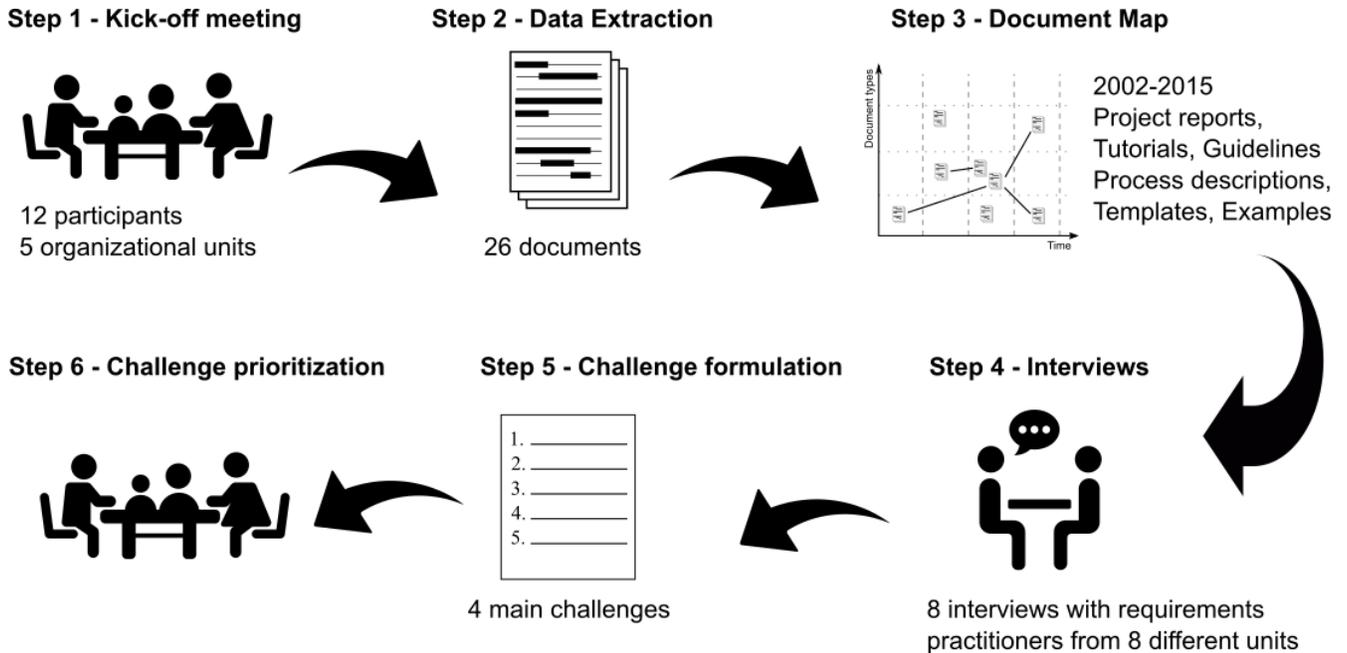

Figure 2: Process Improvement Archaeology (PIA)

# Process Improvement Archaeology

The goal of Process Improvement Archaeology (PIA) is to learn from past improvement initiatives and their outcomes, and identify, based on current data, challenges and improvement opportunities. We developed this method, which is inspired by our previous work on post-mortem analysis [13], to cater for STA's de-centralized improvement strategy where independent initiatives work toward a common goal. This lack of central organization allows for maximum flexibility in the organizations' units to plan improvements, makes it however also more difficult to coordinate and benefit from synergetic effects. Figure 2 illustrates the six PIA steps. We select the participants of the kick-off meeting in Step 1 based on their current involvement in improvement initiatives related to requirements engineering and management. It is important here to identify a diverse set of participants from different organizational units. We elicit from the participants a starting set of documentation produced in past investigations. Similar to literature reviews with snowball sampling [14], the goal is to collect a diverse set of documentation, from as many different sources and authors as possible. In Step 2 we analyze the documentation, extracting the type of the artifact, creation date, authors, purpose and outcome of the investigation, and references to other investigations not included in the original starting set, which we in turn look-up and analyze too. Based on the extracted information, we create in Step 3 a document map that helps to understand relationships, or the lack thereof, between investigations. The structured representation of the investigations lets one to identify those efforts that match the overall goal of the PIA. For example, if the goal is to better understand why improvement efforts seem not to have any discernable impact, one would select those past initiatives that are not referred by more recent ones. Another strategy could be to group initiatives according to their purpose/outcome and check whether they refer to each other or not; this allows one to identify initiatives that would benefit from a closer collaboration. Finally, "fertile" investigations can be identified by looking at often referred documents. This allows one to identify those initiatives that seem to influence the overall improvement direction in an organization. This analysis allows one to identify central improvement initiatives, their purpose and the involved stakeholders which are then interviewed in Step 4. The concrete interview questions depend on the goal of the PIA. A typical goal would be to understand the challenges of implementing recommendations that were produced in an improvement investigation. Another goal could be to understand why initiatives with similar purposes are not using each others' results and findings. In Step 5, we compile a list of challenges, based on the results from the conducted interviews, and form an interest group. Together with this group, we prioritize and develop plans to address the identified challenges in Step 6.

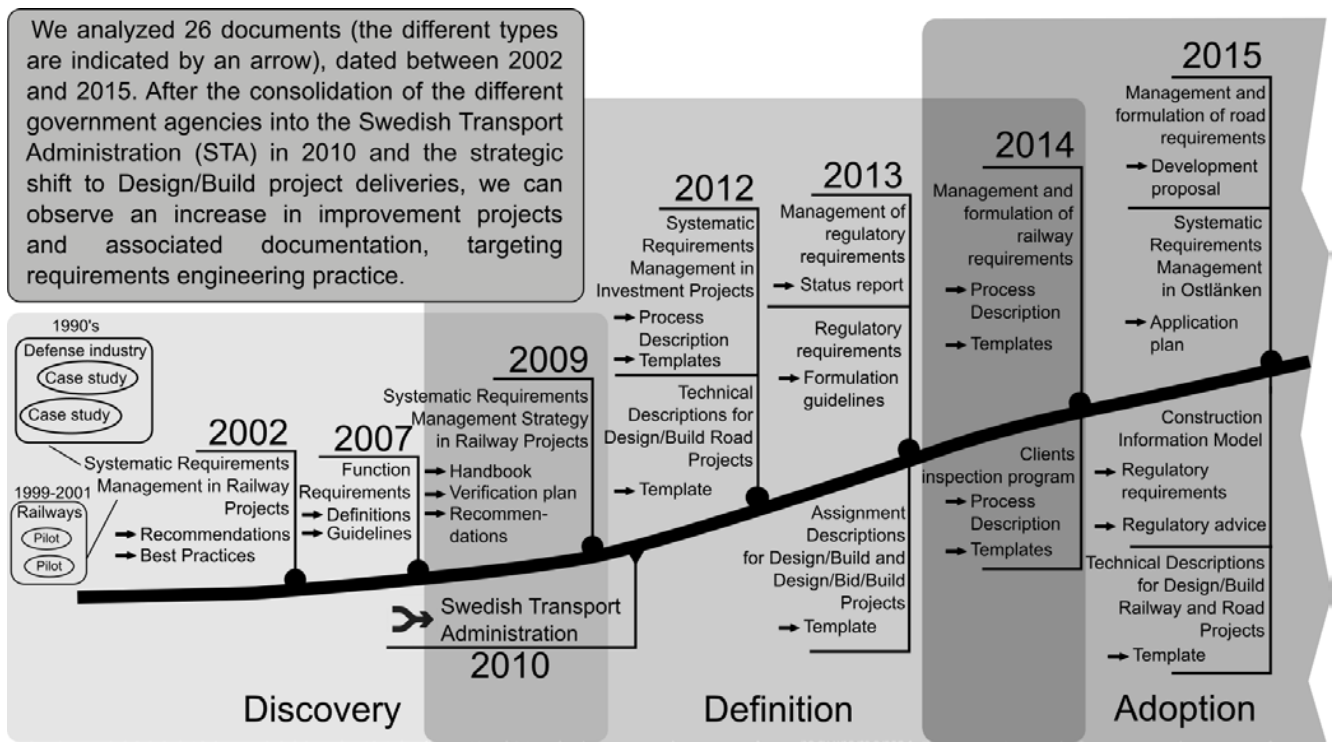

*Figure 3: The development and adoption of Requirements Engineering in STA, illustrated by the various documents produced in internal studies and improvement projects.*

# Learning Requirements Engineering

We now look at the evolution of requirements engineering at STA, the output of Step 3 of the PIA described earlier. Figure 3 illustrates the progression and key documents that can be associated with three overlapping phases in which STA discovered, defined and started adopting systematic requirements management.

## Discovery phase

Investigations in the early 2000's drew experiences from the defense industry on how to perform systematic requirements management, where central concepts such as requirements types and levels, traceability, and verification and validation were analyzed and considered for adoption in railway projects. In pilot studies, customer requirements were broken down into system- and functional requirements in order to define alternative designs with different capacity and cost levels. While the pilot studies concluded that the systematic identification and analysis of requirements was useful, it took several years until these principles were evaluated on a broader basis. The investigation in 2009 reports on the outcome of different strategies to introduce systematic requirements management (SRM) within the investment division of the railway agency. SRM is a process with the following six requirements management concepts: identification, formulation, systematization, acceptance, verification and validation (a seventh concept, change management, was added to the process in 2012). A successful strategy was the introduction of a spreadsheet-based requirements monitoring plan that allowed to follow up the implementation of requirements. Also, an effort was made to adapt existing procedures and templates to raise the quality of the specified requirements. However, few of the changes were eventually adopted. In addition, it turned out to be difficult to find a project where the process could be piloted, mostly due to not being able to identify a project in the starting phase, but also because it was difficult to change the culture and habits in long-running projects.

## Definition phase

This phase is fueled by the 2010 merger of the different transportation agencies into STA and the declared goal to increase the number of Design/Build projects. Until today, processes are defined and templates are produced to support this goal (see Figure 3). In order to promote the new processes and templates, already in late 2010 a competence network was founded, counting as of 2016 more than 130 members from the planning, investment and maintenance division of STA. The SRM process for investment projects was officially approved in 2012. The stated purpose is to:

- ensure that facility requirements are known to all stakeholders when decisions are taken about those requirements
- ensure that no requirements are overlooked during the project
- ensure that the facilities' compliance to the requirements can be evaluated

While the process description is prescriptive in what activities should be performed and when, it does not define *how* the work shall be conducted. The role of the requirements expert is responsible to coordinate the process implementation and adapt it to the projects' needs. Besides developing process descriptions, a large effort was made to collect and synthesize documentation from existing Design/Build projects into templates that capture reusable domain knowledge. For example, the template from 2012, "Technical Descriptions for Design/Build Road Projects" (extended in 2015 to include also railway projects), exhaustively lists and describes all relevant components of a road project in a hierarchical structure. Each component contains generic blueprints for defining scope, functions, requirements on materials, and means to verify the implementation that need to be specified for the particular project.

## Adoption phase

In this latest and current phase we can observe how requirements engineering and management is becoming a central activity in STA (see Figure 3). For example, the clients' inspection program covers how and when to verify deliveries against requirements related to a set of product attributes (e.g. compliance to safety and security regulations and standards). This program would not be effective without the SRM process. Other indications for the adoption of the SRM process is the extension of "Management and formulation of railway requirements" to road projects, and the use of SRM in several large infrastructure projects, one of which is Ostlänken, budgeted for 7 Billion USD with a planned completion in 2028.

# Requirements Engineering – the common ground between engineering disciplines

The document analysis illustrates how STA's decision to increase the number of D/B projects resulted also in an increased interest in fostering requirements engineering competence and expertise. The Systematic Requirements Management (SRM) process, illustrated on the left in Figure 4, is thereby a common theme that evolved in over more than a decade. We decided therefore to focus the PIA interview (see Step 4 in Figure 2) on this process and its actual application. We conducted three interviews with experts communicating the process, who directed us then to five more interviewees with expertise on working with the process.

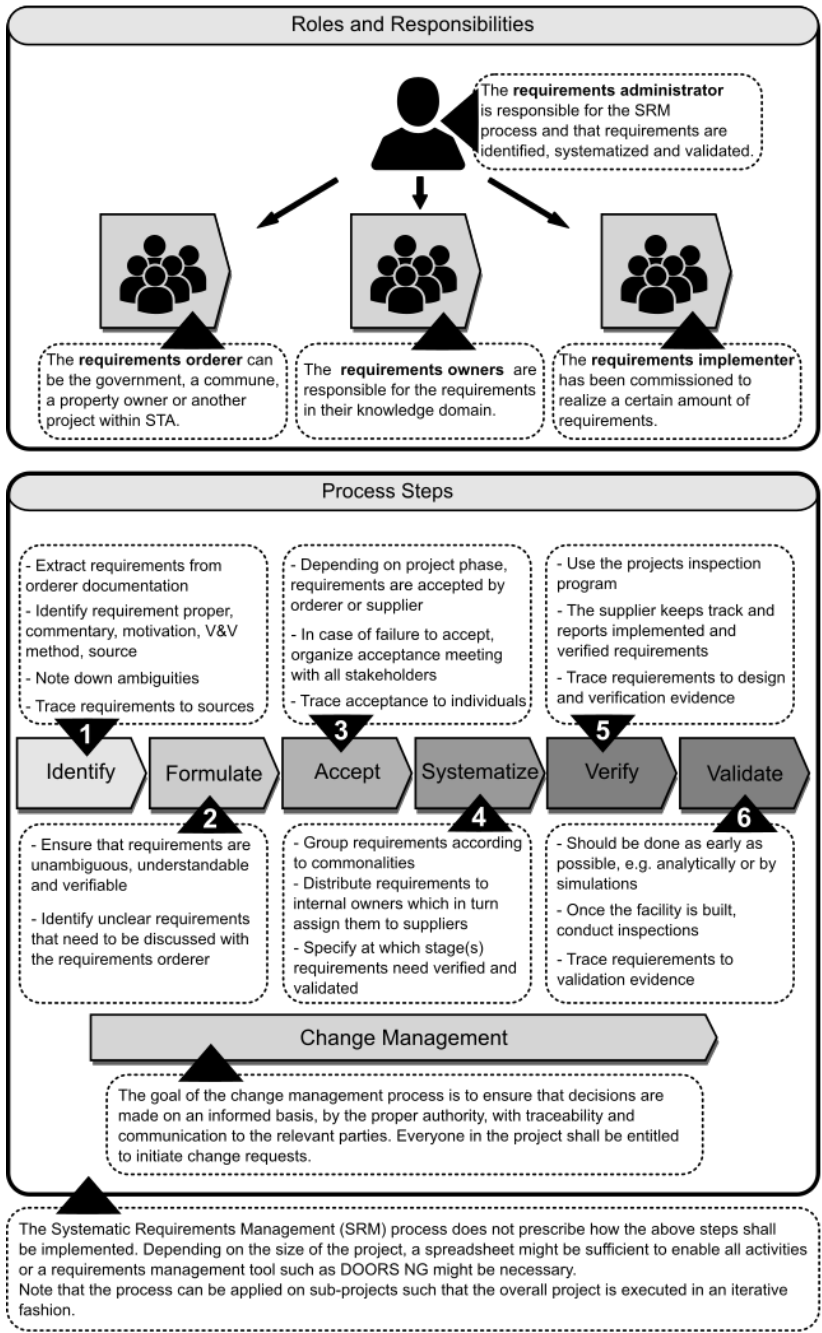

*Figure 4: (Left) Systematic Requirements Management roles and six process steps. (Right) A selection of potential solutions or research reviews from the SE literature for each step.*

The main goal of SRM is to increase the likelihood of building the correct product from the beginning, reducing cost overruns due to rework and loss of information, and increasing planning and development efficiency. In the interviews, we focused on identifying the encountered challenges in implementing SRM, as discussed next, and identify reviews and potential solutions from software engineering to these challenges (referenced in Figure 4, right).

## Identification

STA has a long tradition in designing their facilities together with contractors, following the D/B/B paradigm. The requirements owners (see Figure 4 for a description of roles) have the technical expertise that leads under the D/B paradigm to requirements that tend to specify solutions rather than defining a customer need. One interviewee summarized: "If you choose a solution very early in the process, I don't think you can become a procurement agency. You need to specify your goals, your driving functions and performance requirements."

One strategy to address this challenges is to place the stated requirements on abstraction levels [I1] and work-up (abstract) those that are on a component level, i.e. close to a solution. This would allow STA to negotiate product and function level requirements with the contractors while still benefiting from the in-house technical expertise. In addition, problem framing [I2] would be a useful approach to separate requirements from implementations. The combination and traceability [I4] between requirements on different abstraction level also increases understanding of the requirements, as well as the ability to present only relevant levels of information to appropriate stakeholders – in essence avoiding a flat huge document used by everyone.

Requirements are also elicited from different sources, e.g. the government, property owners, facility users, technical standards, regulation and laws, that may produce overlapping, conflicting and not directly comparable needs. These requirements are large in quantity and heterogeneity, making it very costly to analyze them manually. Recommendation systems [I5] that support domain analysts to identify conflicting and redundant needs can be useful to improve, both in terms of efficiency and quality, the identification of requirements.

## Formulation

The purpose of written requirements is to document and convey information that is understood and used by people with varying domain knowledge. While STA performs seminars and education events to support the requirement orderers, the analysts and requirements editors reported in the interviews that they need to spend effort in reviewing and correcting faults that could be detected earlier. A requirements analyst explained that "there is little support for formulating high quality requirements for those who order them, so we analysts need to improve them". Formulating high quality requirements is costly as many different quality aspects could be considered. It is therefore important to prioritize both quality aspects and critical requirements such that a "good-enough" formulation quality can be achieved for the most central ones. Approaches to evaluate and improve requirements quality [F2,F3] span from computer-based support for correctness, completeness and consistency checking, term recommendations for glossaries, and ambiguity solving to review processes like perspective based reading.

## Acceptance

The specified requirements are the interface between the orderers (stakeholders articulating a need) and implementers (stakeholders fulfilling that need). The interviewees identified requirements acceptance, i.e. where all stakeholder mutually agree on and commit to the stated requirements, as a critical step in the SRM process where cost overruns due to rework can be prevented early in the project: "We have a lot of people whishing things in the project and it is hard to know when we actually have accepted a requirement in the project".

Requirements negotiation [A1] is an important aspect that helps to resolve conflicts among stakeholders, reducing the risk of misunderstandings, making tacit knowledge explicit, and helps finding better solutions. Different conflict resolution strategies can be applied, depending on the goal of the negotiations and the mutual trust of the participating parties. Tools can provide passive support, e.g. by enabling collaboration and communication, or active support, e.g. by facilitating the identification of situations with mutual gain. When negotiating with suppliers, implementation proposals [A2], a technique applied in some projects in STA, can be used to clarify requirements and reduce misunderstandings.

## Systematization

The main goal of this step is to ensure that the requirements are managed such that the relevant stakeholders have access to the information they need to fulfil their duties. The challenge here lies in the fact that projects can last decades while contractors change, rendering handover and information transfer crucial in order to prevent knowledge loss, as one interviewee stated: "We have people come and go in the project and loose knowledge".

In global software engineering, strategies to geographically distribute knowledge could be of great value in this context, where project duration and stakeholder turnover are the major barriers against knowledge conservation [S1]. The invested effort to systematically manage requirements over an extended period of time incites also requirements reuse, both within but also among projects. To enable requirements reuse [S2], approaches to formulate requirements independently from solutions (if it makes economical and technical sense) are needed to not limit innovation and progress.

## Verification & Validation

In the SRM process, verification refers to the assessment whether the design fulfills the specified requirements: "A requirement is something that can be ultimately measured somehow in the final product, i.e. it can be concluded whether the requirement was fulfilled or not". Requirements verifiability is therefore of essence since it is the contractor's responsibility to produce evidence of fulfillment. Quality aspects that contribute to the verifiability of requirements are their design independence, traceability, unambiguousness, understandability, completeness and consistency [V1]. Early focus on improving these quality aspects can reduce the amount of rework, in particular when the product is validated. One possible approach to detect low quality requirements early is to utilize requirements error and source taxonomies [V2] that could help to design focused review processes.

# What did we learn and what's next?

The Process Improvement Archaeology (PIA) turned out to be a useful and lightweight tool that helped us understanding STA's efforts to establish their requirements engineering process and their current challenges. Learning requirements engineering at STA was by no means a straightforward endeavor: an initial discovery phase was followed by a definition phase (fueled by a strategic shift to prioritize the D/B project delivery approach) and the current adoption phase, spanning in total over 15 years. When studying STA's efforts, we were surprised by the parallels to the software industry but also by the long time-frames it takes to establish new practices, likely related to the project cycles ranging from 3 to 20 years. In software engineering, product and process ideas can be developed and validated in much quicker cycles, allowing for an accelerated organizational learning compared to large infrastructure endeavors. Our goal is therefore to support the Systematic Requirements Management process in STA with proven techniques from software engineering, applying them in a new context, and advancing both systems and software engineering. The two directions we are focusing on are related to the identification and formulation of requirements.

First, to utilize abstraction to enable both feature level (what) and component level (how) requirements to be used in combination. This gives the ability of reuse on higher levels, and specifics on lower levels of specification – in essence decoupling the purpose/needs from the solution, while still maintaining technical control. This was successfully realized utilizing methods like the requirements abstraction model in industry.

Second, it is crucial to formulate requirements such that they are fit for purpose and support the respective stakeholders in their tasks. Requirements quality needs to be tuned to be "good enough" while being economically reasonable. Hence, deciding on which quality aspects to improve is equally important as to prioritize requirements that are risky or costly to rework in order to find an acceptable cost/benefit ratio. Such a decision framework would be both beneficial for construction as well as software engineers.

# Biographies

**Michael Unterkalmsteiner** received the BSc degree in applied computer science from Free University of Bozen-Bolzano in 2007, and the MSc and PhD degrees in software engineering from Blekinge Institute of Technology (BTH) in 2009 and 2015, respectively. He is a senior lecturer at BTH. His research interests include software repository mining, software measurement and testing, process improvement, and requirements engineering. He is a member of the IEEE. For more information or contact: www.lmsteiner.com.

**Tony Gorschek** is a Professor of Software Engineering at Blekinge Institute of Technology - where he works as a research scientist in close collaboration with industrial partners. Dr. Gorschek has over fifteen years industrial experience as a CTO, senior executive consultant and engineer, but also as chief architect and product manager. In addition he is a serial entrepreneur – with five startups in fields ranging from logistics to internet based services and algorithmic stock trading. At present he works as a research leader and in several research projects developing scalable, efficient and effective solutions in the areas of Requirements Engineering, Product Management, Value based product development, and Real Agile™ and Lean product development and evolution. For more information or contact: www.gorschek.com

# Tweets

- Process Improvement Archaeology is a lightweight tool to study and learn from past improvement initiatives.

- New practices are only slowly adopted in organizations developing large-scale infrastructure projects

- Learning Requirements Engineering in the Swedish Transport Agency had phases of discovery, definition and adoption

- Requirements challenges in Systems and Software Engineering are similar and can be tackled with common solutions